\documentclass[prd,twocolumn,floatfix,superscriptaddress]{revtex4}

\usepackage{natbib} 
\usepackage{bm}

\usepackage{graphicx}
\usepackage{amsmath}
\usepackage{amssymb}
\usepackage{hyperref}

\newcommand{\LCDM}{$\Lambda\mathrm{CDM}$}
\newcommand{\fr}{$f(R)$}
\renewcommand{\l}{\ell}

\begin{document}

\title{Projected Constraints on Modified Gravity Cosmologies\\ from 21\,cm Intensity Mapping}
\author{Kiyoshi Wesley Masui}
\email{kiyo@cita.utoronto.ca}
\affiliation{Canadian Institute for Theoretical Astrophysics, 60 St.~George Street
Toronto, Ontario, M5S 3H8, Canada}
\affiliation{Department of Physics, University of Toronto, 60 St.~George Street
Toronto, Ontario, M5S 1A7, Canada}

\author{Fabian Schmidt}
\email{fabians@caltech.edu}
\affiliation{Theoretical Astrophysics, California Institute of Technology M/C 35
0-17,
Pasadena, California  91125-0001, USA}

\author{Ue-Li Pen}
\email{pen@cita.utoronto.ca}
\affiliation{Canadian Institute for Theoretical Astrophysics, 60 St.~George Street
Toronto, Ontario, M5S 3H8, Canada}

\author{Patrick McDonald}
\email{pmcdonal@cita.utoronto.ca}
\affiliation{Canadian Institute for Theoretical Astrophysics, 60 St.~George Street
Toronto, Ontario, M5S 3H8, Canada}

\date{February 9, 2010}

\begin{abstract}

We present projected constraints on modified gravity models from
the observational technique known as 21\,cm intensity mapping, where cosmic
structure is detected without resolving individual galaxies.  The resulting
map is sensitive to both BAO and weak lensing, two of the most powerful 
cosmological probes.  It is found that a
$200\,\mathrm{m}\times200\,\mathrm{m}$ cylindrical telescope, sensitive
out to $z=2.5$, would be able to
distinguish DGP from most dark energy models, and constrain the Hu \& Sawicki 
\fr{} model to $|f_{R0}|<9 \times 10^{-6}$ at 95\% confidence.  The latter constraint makes
extensive
use of the lensing spectrum in the nonlinear regime.  These results show
that 21\,cm intensity mapping is not only sensitive to modifications of 
the 
standard model's expansion history, but also to structure growth.
This makes intensity mapping
a powerful and economical technique, achievable on much shorter time scales
than optical experiments that would probe the same era.

\end{abstract}

\maketitle

\section{Introduction}
	\label{s:intro}

	One of the greatest open questions in cosmology is the cause of the
        observed late 
	time acceleration of the universe.   Within the context of normal 
	gravity described by Einstein's General Relativity, this phenomena 
	can only be explained by an exotic form of matter with negative 
	pressure.  Another possible explanation is that on cosmological scales, 
	General Relativity fails and must be replaced by some theory of 
	modified gravity.

	Several approaches have been proposed to modify gravity at late times
	to explain the apparent acceleration of the universe.  The challenge in
	these modifications is to preserve successful predictions of the CMB at
	$z\approx1000$, and also the precision tests at the present epoch
	in the solar system.

	A generic class of theories operates with the Chameleon effect, where at
	sufficiently high densities General Relativity (GR) is restored, thus applying
	both in the solar system and the early universe.  To further understand
	the nature of gravity would require probing gravity on cosmological
	scales.  Large scales means large volume, requiring large fractions
	of the sky.  Gravity can be probed by gravitational lensing,
	which measures geodesics and thus the gravitational curvature of space,
        and is a sensitive probe of the growth of structure in the Universe
        \cite{KnoxSongTyson,JainZhang,Tsujikawa2008,Schmidt:2008hc}.

	In working out predictions for cosmology, the theoretical challenge 
        posed by these theories are the nonlinear mechanisms in each model, necessary
        in order to restore Einstein Gravity locally to satisfy Solar System
        constraints.  We present
	quantitative results from nonlinear calculations for a specific \fr{}
	model, and forecasted constraints for future 21\,cm experiments.

	An upcoming class of experiments propose the observation of the 
	21\,cm spectral line at low resolution over a large fraction of the
	sky and large range of redshifts \cite{Peterson:2009ka}.  Large scale
	structure is detected in three dimensions without the detection of 
	individual galaxies.  This process is referred to as
	\emph{21\,cm intensity mapping}.  These experiments are sensitive to
	structures at a redshift range that is observationally difficult to observe for
	ground-based optical experiments due to a lack of spectral lines.  Yet
	these experiments are extremely economical \cite{Seo:2009fq} since they
	only require limited resolution and no moving parts.
	
	Intensity mapping is sensitive to both the Baryon Acoustic Oscillations
	(BAO) and to weak lensing, two of the most powerful observational 
	methods to determine cosmological parameters.
	It has been shown that BAO detections from
	21\,cm intensity mapping are powerful probes of 
	dark energy, comparing favourably with Dark Energy Task Force
	Stage IV projects within the figure of merit framework
	\cite{Chang:2007xk,Albrecht:2006um}.

	In this paper we present projected constraints on modified gravity models 
	from 21\,cm intensity mapping.  In Section \ref{s:mgm} we describe the 
	modified gravity models considered.  In Section \ref{s:obs} we discuss the
	observational signatures accessible to 21\,cm intensity mapping, and
	calculate the effects of modified gravity on these signatures.  In Section
	\ref{s:results} we present statistical analysis and results and we conclude
	in Section \ref{s:discuss}.
	
	We assume a fiducial \LCDM{} cosmology with WMAP5 cosmological parameters: 
	$\Omega_m = 0.258$, $\Omega_b = 0.0441$, $\Omega_\Lambda = 0.742$, 
	$h = 0.719$, $n_s=0.963$ and $\log_{10}A_s = -8.65$.  We will follow the
	convention that $\omega_x \equiv h^2\Omega_x$.

\section{Modified Gravity Models}
	\label{s:mgm}

	Here we describe some popular modified gravity models for which 
	projected constraints will later be derived.  Throughout we will use units 
	in which $G=c=\hbar=1$ and will be using a metric with mostly negative 
    signature: $(+, -, -, -)$.
	
	\subsection{f(R) Models}
		\label{s:mgm:fr}

		In the \fr{} paradigm, modifications to gravity are introduced 
		by changing the standard Einstein-Hilbert action, which is linear
		in $R$, the Ricci scalar.  The modifications are made by adding an 
		additional non linear function of $R$ 
        \cite{Starobinsky:1980te,gr-qc/0201033,carroll04a}
		\begin{equation}
			S = \int d^4x\sqrt{-g}\left[ \frac{R+f(R)}{16\pi} 
				+ \mathcal{L}_m \right]
			\label{e:mgm:modEHact}
		\end{equation}
		where $\mathcal{L}_m$ is the matter Lagrangian. 
		See \cite{Sotiriou:2008rp} for a comprehensive review of \fr{}
		theories of gravity.

		The choice of the function \fr{} is arbitrary, but in practice it 
		is highly constrained by precise solar system and cosmological 
		constraints, as well as stability criteria
		\cite{Nojiri:2003ft,Sawicki:2007tf} (see below).  In this paper, 
		we choose parameterizations of \fr{} such that it 
		asymptotes to a constant for
		a certain choice of parameters and thus approaches the fiducial
		\LCDM{}.

        In general, \fr{} models have enough freedom to mimic exactly
        the \LCDM{} expansion history and yet still impose a
        significant modification to gravity \cite{Nojiri:2006gh,Song:2007da}.
        As such probes of the expansion history are less constraining than 
        probes of structure growth, which will be evident in the constraints
        presented in later sections.

		Variation of the above action yields the modified Einstein Equations
		\begin{equation}
			G_{\mu\nu}+f_R R_{\mu\nu} - \left(\frac{f}{2}-\Box f_R \right) g_{\mu\nu}-\nabla_{\mu}\nabla_{\nu} f_R = 8 \pi\: T_{\mu\nu}
			\label{e:mgm:modEE}
		\end{equation}
		where $f_R\equiv df(R)/dR$, a convention that will be used 
        throughout.
        \fr{} gravity is equivalent to a scalar-tensor theory 
        \cite{Nojiri:2003ft,Chiba} with
		the scalar field $f_R$ having a mass and potential determined by the
        functional form of \fr{}. The 
		field has a Compton wavelength given by its inverse mass
                \begin{equation}
                  \lambda_C = \frac{1}{ m_{f_R} } = \sqrt{3 f_{RR}}.
                    \label{e:lC}
                \end{equation}   
The main criterion for stability of the $f(R)$ model is that the mass squared
of the $f_R$ field is positive, i.e. $f_{RR} > 0$.  In most cases, this simply corresponds to
a sign choice for the field $f_R$ (specifically for the model we consider
below, $f_{R0}$ is constrained to be less than 0).

		On scales smaller than $\lambda_C$, gravitational forces are
		enhanced by $4/3$, while they reduce to unmodified gravity on
		larger scales. The reach of the modified forces $\lambda_C$
		generically leads to a scale-dependent growth in \fr{} models.

        While the dynamics are significantly changed in \fr{},
        the relation between matter and the lensing potential
        is unchanged up to a rescaling of the
        gravitational constant by the linear contribution in $f$.
        The fractional change is of order
        the background field value $\overline{f_R}\equiv f_R(\bar{R}) \ll 1$
        where $\bar{R}$ is the background curvature scalar.  

		Proceeding further requires a choice of the functional form for $f$.
		A functional form is considered which is
        representative of many
		other cases. 

			Hu and Sawicki proposed a simple functional form for \fr{}
			\cite{Hu:2007nk}, which can be written as
			\begin{equation}
				f(R)= - R_0\frac{c_1 (R/R_0)^n}{c_2 (R/R_0)^n +1},
				\label{e:mgm:fr}
			\end{equation}
            where we have used the value of the scalar curvature in the
			background today, $R_0\equiv \bar{R}|_{z=0}$ for convenience.
			This three parameter model passes all stability criteria for
			positive $n$, $c_1$ and $c_2$.  One parameter can be fixed by
			demanding the
			expansion history to be close (within observational limits)
            to \LCDM{}. In this case, Equation
			\ref{e:mgm:fr} can be conveniently reparametrized and approximated
			by
			\begin{equation}
				f(R) \approx  -2\Lambda - 
					\frac{f_{R0}R_0}{n}\left(\frac{R_0}{R}\right)^n.
				\label{e:mgm:frred}
			\end{equation}
			Here $\Lambda$ and $f_{R0}$---the value of the $f_R$ field in the
			background today---have been used to parameterize the function in
			lieu of $c_1$ and $c_2$.  This approximation is valid as long as
			$|f_{R0}| \ll 1$, which is necessary to satisfy current
			observational
			constraints \cite{Hu:2007nk,fRcluster}. While $\Lambda$ is
			conceptually different than vacuum energy, it is mathematically
			identical and will thus be absorbed into the right hand side
			of the Friedmann equation and parameterized by $\Omega_\Lambda$.
            In quoting constraints, we will marginalize over this parameter
            as it is of no use in identifying signatures of modified gravity.
            The parameter $f_{R0}$ can be though of as controlling the strength 
            of modifications to gravity today, while higher $n$ pushes these
            modifications to later times. The effects of changing these
            parameters are discussed
            in greater detail in \cite{Hu:2007nk}.

            Allowed \fr{} models exhibit the
            so-called chameleon mechanism: the $f_R$ field becomes
            very massive in dense environments and effectively
            decouples from matter.  This effect is active whenever
	    the Newtonian potential is of order the background $f_R$ field.  
	    Since cosmological potential wells are typically of order
	    $10^{-5}$ for massive halos, the chameleon effect becomes
	    important if $|\overline{f_R}| \lesssim 10^{-5}$.  If the
	    background field is $\sim 10^{-7}$ or smaller, a large fraction
	    of the collapsed structures in the universe are chameleon-screened,
	    so that the model becomes observationally indistinguishable from
	    $\Lambda$CDM.
	    
	    Since the chameleon effect will affect the formation of structure,
            standard fitting formulas based on ordinary GR simulations,
            such as those mapping the linear to the nonlinear power spectrum,
            cannot be used for these models. 
            Recently, however, self-consistent N-body simulations
            of \fr{} gravity have been performed which include
            the chameleon mechanism \cite{HPMpaper,HPMpaperII,
            HPMhalopaper}.  We will use the simulation
            results for forecasts of weak lensing in the nonlinear
            regime below.

			It should be noted that \fr{} models are not without
			difficulties.  
			In particular, an open issue is the problem of potential 
			unprotected singularities
			\cite{Abdalla:2004sw,Frolov:2008uf,Nojiri:2008fk}.

	\subsection{DGP Braneworld}
		\label{s:mgm:dgp}

		A theory of gravity proposed by Dvali, Gabadadze and Porrati (DGP)
		assumes that our four dimensional universe sits on a brane in five
		dimensional Minkowski space \cite{Dvali:2000hr}.  On small scales
		gravity is four dimensional but, on larger scales it becomes fully
		five dimensional.  Here we parameterize DGP by $r_c$, the scale at
		which gravity crosses over in dimensionality.  The DGP model has two
		branches depending on the embedding of the brane in 5D space. In
		the self-accelerating branch, the universe accelerates without need
		for a cosmological constant if $r_c \sim 1/H_0$
		\cite{Deffayet:2000uy, Deffayet:2001pu}. In this branch, assuming a 
        spatially flat Universe for now, the modified
		Friedmann equation is given by
		\begin{equation}
			H^2 - \frac{H}{r_c} = \frac{8\pi}{3} \bar{\rho},
			\label{e:HDGP}
		\end{equation}
		which clearly differs from \LCDM{}. Thus, in contrast to the other 
		models considered here, DGP without a cosmological constant does
		not reduce to \LCDM{} and it is possible to completely rule
		out this scenario (where the others can only be constrained).  In fact
		DGP (without a cosmological constant) has been shown to be in
                conflict with current data
		\cite{Fang:2008kc}.  It is presented here largely for illustrative
		purposes.

		On scales much smaller than $r_c$, gravity is four-dimensional but
		not GR. On these scales, DGP can be described as an effective
		scalar-tensor theory \cite{KoyamaMaartens,KoyamaSilva,Scoccimarro09}.
		The massless scalar field, the brane-bending mode, is repulsive in
		the self-accelerating branch of DGP. Hence, structure formation is
		slowed in DGP when compared to an effective smooth Dark Energy model
		with the same expansion history.  While the growth of structure
        is thus modified in DGP even on scales much smaller than $r_c$,
        gravitational lensing is unchanged.  In other words, the relation
        between matter overdensities and the lensing potential is the
        same as in GR \cite{LueEtal04}.

        As in \fr{}, the DGP model contains a nonlinear mechanism
        to restore GR locally.  This Vainshtein mechanism is due to 
        self-interactions of the scalar brane-bending mode which generally become
        important as soon as the density field becomes of order unity.
        In the Vainshtein regime, second derivatives of the field
        saturate, and thus modified gravity effects are highly
        suppressed in high-density regions \cite{LueEtal04,KoyamaSilva,DGPMpaper}.
        We will only consider linear predictions for the DGP model here.

\section{Observational Signatures}
	\label{s:obs}

	In this Section we describe the observational signatures available to 
	21\,cm intensity mapping.  We also give details on calculating the
	observables within modified gravity models.  We consider two types of
	measurements: the Baryon Acoustic Oscillations and weak gravitational 
	lensing.

    For the fiducial survey, we assume a $200\,\mathrm{m}\times200\,\mathrm{m}$ 
    cylindrical telescope, as in \cite{Chang:2007xk}.
    We will also present limited results for a $100\,\mathrm{m}\times100\,\mathrm{m}$
    cylindrical telescope to illustrate effects of reduced resolution and 
    collecting area on the results.
    This latter case is representative of first generation 
    projects \cite{Seo:2009fq}. In the 200\,m case we assume 4000 receivers, and 
    in the 100m case 1000 receivers. We assume either telescope covers 
    15000 sq. deg.  over 4 years.
    We assume neutral hydrogen fraction and the bias remain constant with
    $\Omega_{\rm HI}=0.0005$ today and $b=1$.  The 
    object number density is assumed to be $\bar{n}=0.03$ per cubic 
    $h^{-1}\rm{Mpc}$
    (effectively no
    shot-noise, as should be the case in practice \cite{Chang:2007xk}).

	\subsection{Baryonic acoustic oscillation expansion history test}
    	\label{s:obs:exphist}

		Acoustic oscillations in the primordial photon-baryon plasma
		have ubiquitously left a distinctive imprint in the distribution of
		matter in the universe today.  This process is understood from
		first principles and gives a clean length scale in the universe's 
		large scale structure, largely free of systematic uncertainties and 
		calibrations.  This can be used to measure the global cosmological 
		expansion history through the angular diameter distance, $d_A$, and 
        Hubble parameter, $H$, vs 
		redshift relation. The detailed expansion and acceleration will 
		differ between
		pure cosmological constant and modified gravity models.

We use essentially the method of \cite{2007ApJ...665...14S} for estimating
distance errors obtainable from a BAO measurement, including 50\% 
reconstruction of nonlinear degradation of the BAO feature.
We assume the frequency range corresponding to $z<2.5$ is covered (the lower
$z$ end should be covered by equivalent galaxy redshift surveys if not a 21cm
survey).
For the sky area and redshift range surveyed, the 200m telescope is nearly 
equivalent to a perfect BAO measurement. The limited resolution and collecting
area of the 100m telescope substantially degrades the measurement at the 
high-$z$ end.

The expansion history for modified gravity models can be calculated
in an analogous way to that in General Relativity.  The Friedmann Equation
in DGP, Equation~\ref{e:HDGP} can be written as
		\begin{equation}
			H^2 = -\frac{k}{a^2} + \left( \frac{1}{2r_c} + 
              \sqrt{\frac{1}{4{r_c}^2} + 
              \frac{8 \pi \bar{\rho}}{3}} \right)^2
			\label{e:obs:DGPfried}
		\end{equation}
		where $k$ is the curvature, and $r_c$ is the crossover scale.
        It is convenient to introduce the parameter $\omega_{rc} \equiv 1/4r_c^2$
        which stands in for $r_c$.
        This equation can be solved numerically
		to calculate the observable quantities.

        We now calculate the expansion history in the HS \fr{} model using
        a perturbative framework which is well suited for calculating 
        constraints on $f_{R0}$.  Working in the conformal gauge and mostly
        negative signature, we start with the modified Einstein's Equation 
			\eqref{e:mgm:modEE}.  At zeroth order the left hand side of the 
			00 component contains the modified Friedmann equation
			\begin{equation}
				H^2 = \frac{8 \pi \bar{\rho}}{3} + 
                  f_{R0}g_n(a,\dot{a},\ddot{a},\dddot{a})
				\label{e:obs:modfried}
			\end{equation}
			where $\bar{\rho}$ is the average density 
            (including contributions from $\Lambda$), the over-dot 
            represents a conformal derivative and
			\begin{eqnarray}
				g_n & \equiv &  \frac{-1}{f_{R0}a^2}
                  \bigg[ \frac{(\bar{f}+2\Lambda)a^2}{6}+
                  \overline{f_R}\left(\frac{\dot{a}^2}{a^2} -
                  \frac{\ddot{a}}{a}\right) \nonumber \\*
				& & + 6\overline{f_{RR}}\left( \frac{\dddot{a} 
                \dot{a}}{a^4}-\frac{3\ddot{a}\dot{a}^2}{a^5}\right) \bigg].
				\label{e:obs:gfunc}
			\end{eqnarray}
			For verifiability we quote
			\begin{equation}
				g_1 = \frac{a^2 R_0^2(2a\ddot{a}^2-7\ddot{a}\dot{a}^2+
                  2\dddot{a}\dot{a}a)}{36\ddot{a}^3}.
				\label{e:obs:g1}
			\end{equation}
			Evaluating Equation \ref{e:obs:modfried} at the present 
			epoch yields the modified version of the standard constraint
			\begin{equation}
				h^2 = \omega_m + \omega_r + \omega_k + 
					\omega_{\Lambda} + f_{R0}g_{n0}.
				\label{e:obs:constraint}
			\end{equation}

			 Note that the modified version of the Friedmann Equation is 
			 third order instead of first order, however, it has been shown
			 that the expansion history stably approaches that of \LCDM{} 
			 for vanishing $f_{R0}$ \cite{Hu:2007nk}.  For observationally 
			 allowed cosmologies 
			 $f_{R0} \ll 1$ we expand
			 \begin{equation}
			 	f_{R0}g_n = f_{R0}g_n(\tilde{a},\dot{\tilde{a}},
                  \ddot{\tilde{a}},\dddot{\tilde{a}}) + O(f_{R0}^2)
				\label{e:obs:approx}
			\end{equation}
			where $\tilde{a}$ is the solution to the \emph{standard}
			GR Friedmann equation.

			By using Equation \ref{e:obs:approx} in Equation 
			\ref{e:obs:modfried} and keeping only terms linear in $f_{R0}$, 
			the expansion history can be calculated in the regular way, 
			along with the observable quantities $d_A(z)$ and $H(z)$.  For 
			small $f_{R0}$ this agrees well with the calculation in 
			\cite{Hu:2007nk} where the full third order differential 
			equation was integrated
			
			In calculating the Fisher Matrix, this treatment is exact
			because the Fisher Matrix depends only on the first derivative
			of the observables with respect to the model parameters, evaluated
			at the fiducial model.
	
	\subsection{Weak Lensing}
		\label{s:obs:wl}

		A second class of observables measures the spatial perturbations
		in the gravitational metric.
		Modified gravity will change the strength of gravity on large
		scales
        and thus modify the
		growth of cosmological structure.  Weak gravitational
		lensing, the gravitational bending of source light by intervening
		matter, is a probe of this
		effect.

		Weak lensing measures the distortion of background structures as their
		light propagates to us.  Here, the background structure is the
        21\,cm emission from unresolved sources.
        While light rays are deflected by gravitational forces,
		this deflection is not directly observable, since we don't know the
        primary unlensed 21\,cm sky.
		However, weak lensing will induce correlations in the measured 
		21\,cm background, 
		since neighbouring rays pass through common lens planes.  
        While the deflection angles themselves are small
        (of order arcseconds)
        the deflections are coherent over scales of arcminutes.
        In this way, the lensing signal can be extracted statistically
        using quadratic estimators \cite{Lu:2009je}.
        Given the smallness of the lensing effect, a high resolution
        (high equivalent number density of ``sources'') is necessary
        to detect the effect.

		The weak lensing observable that is predicted by theory is the
		power spectrum of the convergence $\kappa$.  It is given by
\begin{equation}             
C^{\kappa\kappa}(\l) = \left(\frac{3}{2}\Omega_m\:H_0^2 \right)^2 \int_0^{\chi_s}
\frac{d\chi}{\chi} \frac{W_L(\chi)^2}{\chi \: a^2(\chi)}\epsilon^2(\chi)  P(\l/\chi;\chi)
\label{e:Ckappa}
\end{equation}             
		where $\chi$ denotes comoving distances, $P(k,\chi)$ is the
        (linear or nonlinear) matter power spectrum at the given 
		redshift, and we have assumed flat space. The lensing weight 
		function $W_L(\chi)$ is given by:
\begin{equation}             
W_{L}(\chi) = \int_{z(\chi)}^{\infty} dz_s \frac{\chi}{\chi(z_s)}(\chi(z_s)-\chi)\: \frac{dN}{dz}(z_s).
\end{equation}             
Here, $dN/dz$ is the redshift distribution of source galaxies,
normalized to unity. The factor $\epsilon(\chi)$ in Equation~\ref{e:Ckappa}
encodes possible modifications to the Poisson equation relating the
lensing potential to matter (Section~\ref{s:mgm}). In \fr{}, it is given by 
$\epsilon(\chi) = (1 + \overline{f_R}(\chi))^{-1}$,
while $\epsilon=1$ for GR as well as DGP. Note that for
viable \fr{} models, $\epsilon-1 \lesssim 0.01$, so the effect 
of $\epsilon$ on
the lensing power spectra is very small.

The CAMB Sources module \cite{Lewis:1999bs,Lewis:2007kz} was
used to calculate 
the lensing convergence power spectrum in flat \LCDM{} models.
The HALOFIT \cite{halofit} interface for CAMB
was used for calculations that include lensing at nonlinear 
scales.  

For the modified gravity models in the linear regime, the convergence power 
spectra were calculated using the Parametrized Post-Friedmann (PPF) approach
\cite{HuSawickiPPF} as in \cite{Schmidt:2008hc}. Briefly, the PPF approach
uses
an interpolation between super-horizon scales and the quasi-static limit. On 
super-horizon scales ($k \ll aH$), specifying the background expansion history, 
together with a relation between the two metric potentials, already determines 
the evolution of metric and density perturbations. On small scales ($k\gg aH$),
time derivatives in the equations for the metric perturbations can be neglected
with respect to spatial derivatives, leading to a modified Poisson equation
for the metric potentials.  The PPF approach uses a simple interpolation
scheme between these limits, with a few fitting parameters adjusted to match the 
full calculations \cite{HuSawickiPPF}. The full calculations are reproduced to
within a few percent accuracy. We use the transfer function of \cite{EisensteinHu}
to calculate the \LCDM{} power spectrum at an initial redshift of $z_i=40$, were
modified gravity effects are negligible, and evolve forward using the PPF equations.

For the \fr{} model, we also calculate predictions in the nonlinear regime. For these,
we use simulations of the HS model with $n=1$ and $f_{R0}$ values ranging from $10^{-6}$
to $10^{-4}$. We use the deviation $\Delta P(k)/P(k)$ of the
nonlinear matter power spectrum measured in \fr{} simulations from that of 
\LCDM{} simulations with the same initial conditions \cite{HPMpaperII}.
This deviation is measured more precisely than $P(k)$ itself. 
We then spline-interpolate the measurements of $\Delta P(k)/P(k)$ for
$k=0.04 - 3.1\:h/\rm Mpc$ and at scale factors $a=0.1,0.2,...1.0$, and multiply the standard
nonlinear \LCDM{} prediction (HALOFIT) with this value. For values of $k > 3.1\:h/\rm Mpc$,
we simply set $\Delta P(k)=0$. However, for the angular scales and redshifts considered
here ($\l < 600$, see below), such high values of $k$ do not contribute significantly.

One might be concerned that this mixing of methods, for calculating the lensing
spectrum, might artificially exaggerate the effects of modified gravity if these
methods do not agree perfectly.
While the spectra calculated for the fiducial \LCDM{} model differed by up to
a percent between these methods, presumably due to slight differences in the 
transfer function, this should have no effect on the results.  
Any direct comparison between spectra (for example finite difference derivatives)
are made between spectra calculated in the same manner.  Note that the Fisher
Matrix depends only on the first derivative of the observables with respect
to the parameters and no cross derivatives are needed.
		
		The lensing spectra were not calculated for non-flat models, 
		but 
		it is expected that the CMB and BAO are much more sensitive to 
		the
		curvature and as such the lensing spectra are relatively unaffected.
        Formally we are assuming that 
        \[\frac{\sigma_{\omega_k}}{\sigma_{C^{\kappa\kappa}}}\frac{\partial C^{\kappa\kappa}}{\partial \omega_k} \ll 1.\]

		Reconstructing weak lensing from 21\,cm intensity maps
		involves the use of quadratic estimators to
		estimate the convergence and shear fields. The accuracy with which this
		can be done increases with information in the source maps, however, 
		this information saturates at small scales due to nonlinear evolution.
		As such, one cannot improve the lensing measurement indefinitely by
		increasing resolution, and the experiments considered here extract much
		of the available information within the redshift range considered.
		
		The accuracy with which the convergence power spectrum can be 
		reconstructed from 21\,cm intensity maps was derived in 
		\cite{Lu:2009je}, where the effective lensing galaxy density was 
		calculated at redshifts 1.25, 3 and 5 (see Figure 7 and Table 
		2 therein).  
		The effective volume galaxy density was corrected 
		for the finite resolution of the experiment considered here.  
		It was then interpolated, using a piecewise power law, and integrated 
		from redshift 1 to 2.5 to obtain an effective area galaxy density 
		of $n_g/\sigma_e^2 = 0.37 \mathrm{arcmin}^{-2}$. The parameter 
		$\sigma_e^2$ is the variance in the intrinsic galaxy ellipticity,
		which is only used here for comparison with optical weak lensing
		surveys. From the effective galaxy density the error on
		the convergence power is given by
		\begin{equation}
			\Delta C^{\kappa\kappa}(\l) = \sqrt{\frac{2}{(2\l+1)f_{sky}}}
			\left(C^{\kappa\kappa}(\l)+\frac{\sigma_e^2}{n_g}\right)
			\label{e:obs:deltaCl}
		\end{equation}
		where $f_{sky}$ is the fraction of the sky surveyed.
		The galaxy distribution function $dN/dz$ used 
		to calculate the theoretical curves (from Equation \ref{e:Ckappa})
		should follow the effective 
		galaxy density.  Instead for simplicity, a flat step function was 
		used, with this distribution function equal from redshift
		1 to 2.5 and zero 
		elsewhere. While the difference between the these distributions would
have an effect on the lensing spectra, the effect on differences of spectra
when varying parameters is expected to be negligible. 
        Our approximation is also conservative, since the proper distribution
        function is more heavily weighted toward high redshift.  Rays travelling
        from high redshift will be affected by more intervening matter and thus
        experience more lensing.  This would increase the lensing signal, 
        allowing a more precise measurement.

\begin{figure}
			\centerline{\includegraphics[scale=1]{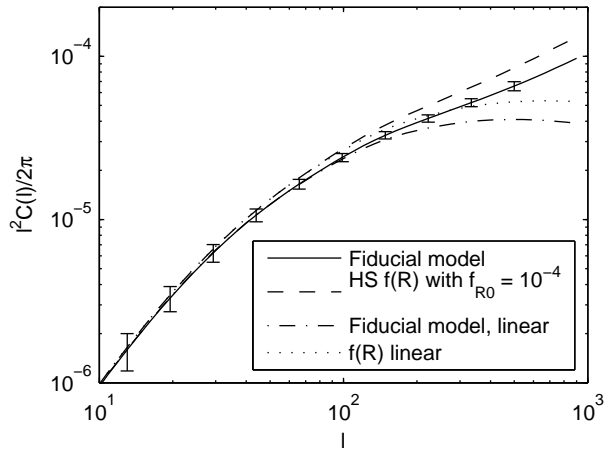}}
			\caption{\label{f:obs:lensing} The Weak lensing convergence 
			power spectra for \LCDM{} and the HS \fr{} model with $n = 1$ and
			$f_{R0} = 10^{-4}$.  Galaxy distribution function is flat between
			z = 1 and z = 2.5.}
		\end{figure}

		Figure \ref{f:obs:lensing} shows the lensing spectra for the fiducial
		cosmology and a modified gravity model, including both linear and 
		nonlinear calculations.  The linear regime is taken to be up to 
		$\l=140$ for projected constraints.  For calculations including 
		weak lensing in the nonlinear regime, $C^{\kappa \kappa}(\l)$ up 
		to $\l=600$ is used for the larger telescope.
        Beyond this scale the model used for lensing 
		error-bars is not considered accurate at the shallowest redshifts
		in the source window \cite{Lu:2009je}.  This cut off coincides with
        the scale at which information in the source structures saturates
        due to non-linear evolution in standard gravity (although it is also
        not far from the resolution limit of the experiment).  We speculate that a 
        similar phenomena would occur in modified gravity and smaller scales 
        are not expected to
        carry significant additional information.  Note that it is the 
        \emph{source} structures in which information saturates.  At smaller 
        scales the lensing spectrum would continue to carry information 
        \cite{Dore:2009jf} if it could be reconstructed.
        For the smaller telescope
        the scale is limited to $\l<425$ by the resolution at the high end
        of the redshift
        window.  If the redshift window were subdivided into narrower bins, 
        it would be possible to use information at scales down to 
        $\l\approx 1000$ in the centre bins as at these redshifts the 
        telescope resolutions are better and structures are less non-linear.
        However, considering tomographic information is beyond the scope of
        this work.
        It is noted that these scales
		are very large by weak lensing standards where optical
		surveys typically make detections down to an $\l$ of order
		$10^5$.
		
	\subsection{External Priors from Planck}
		\label{s:obs:planck}
		While the CMB is not sensitive to the late time effects of modified 
		gravity (except by the integrated Sachs-Wolfe effect), it is 
		invaluable 
		for constraining other parameters and breaking degeneracies.  
		As such,
		projected information from the Planck experiment is included.  The 
		Planck covariance matrix used here is given in
		\cite[Table II]{McDonald:2006qs}.  All late time cosmological 
		parameters (including the curvature) are marginalized over, 
		removing 
		information contained in the ISW effect, and ensuring that 
		sensitivity 
		to \fr{} is entirely from 21\,cm tests below.  The only remaining 
		parameter that is related to the late time expansion is 
		$\theta_s$, the 
		angular size of the sound horizon, which is then used as a 
		constraint on the 
		parameter sets of the modified gravity models.

\section{Results}
	\label{s:results}

	To quantify the projected constraints on \fr{} models, the Fisher matrix
	formalism is employed. The HS \fr{} models
    reduces to the fiducial model for 
	vanishing $f_{R0}$ and any value of $n$.  Thus the Fisher Matrix 
	formalism is used to project constraints on $f_{R0}$ for given values of $n$.
    In the
	case of DGP, which does not reduce to the fiducial model, it is shown
	that a measurement consistent with the fiducial model can not be consistent
	with DGP for any parameter set.  Unless otherwise noted, we account for
	freedom in the full cosmological parameter set: $h$, $\omega_m$, $\omega_b$,
	$\omega_k$, $A_s$ and $n_s$; representing the Hubble parameter; physical
	matter, baryon and curvature densities; amplitude of primordial scalar
	fluctuations and the spectral index; respectively.

	Within the \fr{} models, the fiducial model is a special point in the
	parameter space as there are no modifications to gravity.  As such,
	one cannot in general expect perturbations to observables to be linear in the
	\fr{} parameter $f_{R0}$, an assumption implicit in the 
	Fisher Matrix formalism.  This assumption does seem to hold for 
	the expansion history, where our first order perturbative calculation
	agrees with the full solution to the modified Friedmann Equations 
	calculated in \cite{Hu:2007nk}.  However, this is not the case for
	weak lensing.  For each \fr{} model, the lensing spectrum
	was calculated for several values of $f_{R0}$.  It was observed 
	that enhancements to the lensing power spectrum go as 
    \[C^{\kappa\kappa}(\l)-C^{\kappa\kappa}_{fiducial}(\l) \sim (f_{R0})^{\alpha(\l)}\]
    with $\alpha(\l)$ in the 
    0.5--0.7 range.  This is because the reach of the enhanced forces in \fr{}
    is a power law in $f_{R0}$ following Equation~(\ref{e:lC}), and the
    enhancement of
    the power spectrum for a given mode $k$ roughly scales with the time
    that this mode has been within the reach of the enhanced force. 
    Because of this behaviour, 
    the constraints derived within the Fisher Matrix
	formalism depend on the step size in $f_{R0}$ used for finite
	differences.

	To correct for this, we use a step size that is dependent on the final
	constraint.  The weak lensing Fisher Matrices where calculated for
	$f_{R0}$ step sizes of $10^{-3}$, $10^{-4}$ and $10^{-5}$.
	These were then interpolated---using a power law---such that the ultimate
	step size used for finite differences is roughly the quoted constraint
	on the modified gravity parameter.  For instance when the 95\% confidence
    constraint on $f_{R0}$ is quoted, the step size for finite 
	differences is $\Delta f_{R0} \approx 2\sigma_{f_{R0}}$, where 
    $\sigma_{f_{R0}}$ is calculated from the interpolated Fisher matrix.  This is 
	expected to be valid down to step sizes at the $10^{-6}$ level where
	the chameleon mechanism is important.  As such, for constraints below
	$10^{-5}$ a step size of $10^{-5}$ is always used.  Note that this is 
	conservative because an over sized finite difference step always 
	\emph{underestimates} the derivative of a power law with an power less
	than unity. For constraints above the $10^{-3}$
    level a step size of $10^{-3}$ is used, which is the 
	largest modification to gravity simulated.  These constraints are considered
    unreliable due
	to  these difficulties.  We reiterate this this only affects results that
    include weak lensing information.
    Likelihood contours remain perfect ellipses in this
	procedure (which is clearly inaccurate), however the spacing between
	contours at different confidence levels is altered.

\begin{figure}
		\centerline{\includegraphics[scale=1]{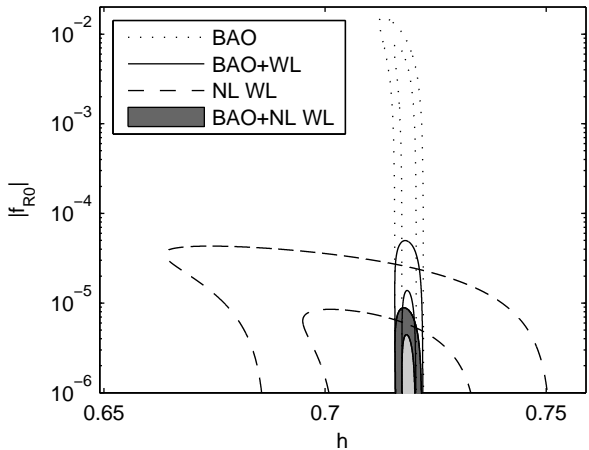}}
		\caption{\label{f:results:FRfisher} Projected constraints on the 
		HS \fr{} model with $n = 1$ using several combinations of observational
		techniques, for a 200\,m telescope. All curves include forecasts for Planck.
        Allowed parameter values are shown in the 
		$f_{R0}-h$ plane at the 68.3\%, and 95.4\% confidence level.
		Results are not shown for ``WL'' which were calculated
		much less accurately (see text).}
	\end{figure}

	Figure \ref{f:results:FRfisher} shows the projected constraints on the HS
	\fr{} model with $n=1$ for various combinations of observational techniques, and a $(200\rm m)^2$ telescope.
	The elements in the lensing fisher matrix associated with the curvature are 
	taken to be zero for the reasons given in Section \ref{s:obs:wl}. While this
	assumption is not conservative, it is expected to be valid, as the angular 
	diameter distance as measured by the BAO is very sensitive to the curvature. 
	In total three \fr{} models were considered: HS with $n=1,2,4$.  
	The results are summarized in Table \ref{t:results:fr}.

	\begin{table}
		\begin{tabular}[c]{|l|c|c|c|c|}
			\hline
			95\% confidence & \multicolumn{4}{|c|}{HS~~$ |f_{R0}|$}  \\
			\cline{2-5}
			upper limits & \multicolumn{1}{|c|}{$n=1$} & \multicolumn{1}{|c|}{NL WL}  & \multicolumn{1}{|c|}{$n=2$} & \multicolumn{1}{|c|}{$n=4$}  \\
			\hline
			BAO & 1.5e-02 & $\sim$ & 1.8e-02 & 3.0e-02 \\ 
WL & 2.3e-03 & 4.3e-05 & 4.0e-03 & 8.6e-03 \\ 
BAO+WL & 5.0e-05 & 8.9e-06 & 9.7e-05 & 4.6e-04 \\ 

			\hline
		\end{tabular}
		\caption{\label{t:results:fr} Projected constraints on \fr{} models
		for various combinations of observational techniques, for a 200\,m telescope.  Constraints
		are the 95\% confidence level upper limits and include forecasts for Planck.
        The non linear results 
		(column marked NL~WL) are for
		the HS model with $n=1$.  Results that make use of weak lensing with
		constraints above $10^{-3}$ are only order of magnitude accurate.
        The linear regime is taken to be $\l < 140$, with the nonlinear constraints
        extending up to $\l=600$.
        }
	\end{table}

	It was found that while weak lensing, in the linear regime,
	is very sensitive to the modifications to gravity, it is only barely 
	capable of constraining \fr{} models without separate information about the
	expansion history.  Even with the inclusion of Planck forecasts, 
	degeneracies with $h$ and $\omega_k$, the mean curvature,
	drastically increase the uncertainties on the modified gravity parameters.
	Indeed these three parameters are more than 95\% correlated (depending on
	the exact model and confidence interval).   This of
	course brings into question the neglect of the $\omega_k$ terms in the weak lensing
	Fisher Matrix.  However it is noted that in these cases, the predicted
    limits on the curvature are $|\omega_k|<0.025$ at 95\% confidence. The current, model
    independent, limits on the curvature using WMAP, SDSS
    and HST data are approximately half this value \cite{Komatsu:2010fb}.
    Our neglect of any direct probes of the expansion history for the Planck+WL
    constraints is clearly unrealistic; however, the constraints illustrate what
    is actually measured by weak lensing.
    In any case these degeneracies are broken once BAO 
	measurements are included, and in this final case the modified gravity
	parameters are correlated with the other parameters by at most 35\%. 
	Also, considering lensing in the nonlinear regime breaks the degeneracy
	to a certain extent.

\begin{figure}
		\centerline{\includegraphics[scale = 1]{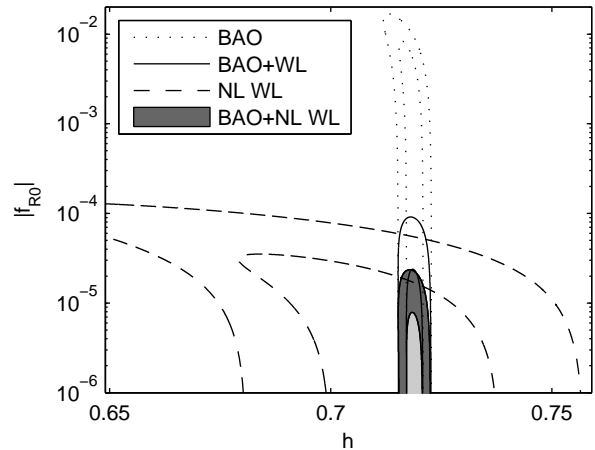}}
		\caption{\label{f:results:FRlowres} Same as Figure \ref{f:results:FRfisher} but for a 
        100\,m cylindrical telescope.}
	\end{figure}

	First generation cylindrical telescopes will likely be smaller than
	the one considered above.  To illustrate the differences in constraining
	ability,
	we now present a few results for a cylindrical radio telescope that is
	$100 \mathrm{m}$ on the side. Reducing the resolution of the experiment
	degrades measurements in a number of ways.  BAO measurements become less
	than ideal in the higher redshift bins.  The smallest scale that can be
	considered for weak lensing drops to about $\l=425$.  A more important
	effect is that the lensing spectra can not be as accurately reconstructed,
	dropping the effective galaxy density down to $n_g/\sigma_e^2 = 0.22$.
	Figure \ref{f:results:FRlowres} shows analogous results to
	\ref{f:results:FRfisher} but for a telescope with half the resolution.

	To show that a set of measurements consistent with the fiducial model
	would be inconsistent with DGP we first fit DGP to the fiducial model's
	CMB and BAO expansion history by minimizing
	\begin{equation}
		\label{e:results:chi2}
		\tilde{\chi}^2 = (\bm{r}_{DGP}-\bm{r}_{fiducial})^T C^{-1}(\bm{r}_{DGP}-\bm{r}_{fiducial})
	\end{equation}
	where $\bm{r}_{DGP}$ and $\bm{r}_{fiducial}$ are vectors of observable
	quantities as calculated in the DGP and fiducial models, and $C$ is the
	covariance matrix.  $\bm{r}$ includes BAO $d_A(z)$ and $H(z)$ as well
	as Planck priors on $\omega_m$ and $\theta_s$.  Note that $\tilde{\chi}^2$
	is not truly chi-squared since $\bm{r}_{fiducial}$ contains fiducial model
	predictions and is not randomly distributed like a real data set.

\begin{figure}
		\centerline{\includegraphics[scale=1]{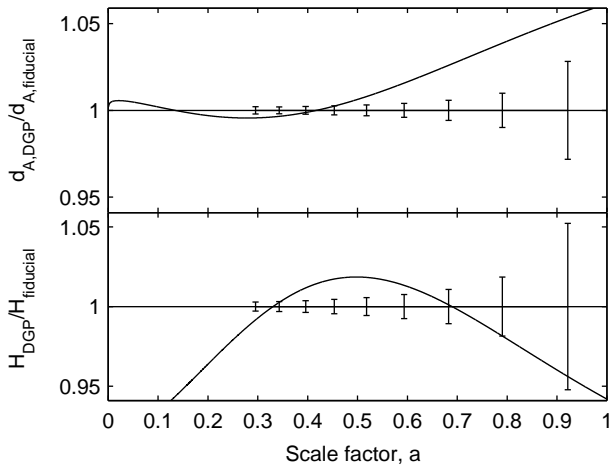}}
		\caption{\label{f:results:DGPBAO} Ratio of the coordinate $d_A(z)$ (top)
		and the Hubble parameter $H(z)$ (bottom) as predicted by the best fit
		DGP model to the fiducial model.  Error bars are from 21\,cm BAO 
		predictions.  Fit includes BAO data available from the 200\,m telescope
        and CMB priors on $\theta_s$ 
		and $\omega_m$.}
	\end{figure}

	Performing the fit yields DGP parameters: $h = 0.677$, 
	$\omega_m = 0.112$, $\omega_k=-0.0086$ and $\omega_{rc} = 0.067$. 
	Figure \ref{f:results:DGPBAO} shows the deviation of $H$ and $d_A$ 
	respectively for the best fit DGP model compared to the fiducial 
	model.  $\tilde{\chi}^2 = 332.8$ for the fit despite there only being 
	16 degrees of freedom, and as such a measurement consistent with the 
	fiducial model would thoroughly rule out DGP.

\begin{figure}
		\centerline{\includegraphics[scale=1]{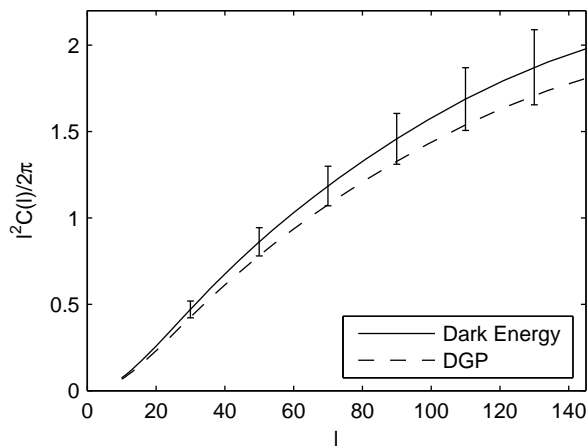}}
		\caption{\label{f:results:DGPWL} Weak lensing spectra in for DGP
		and a smooth dark energy model with the same expansion history. 
		DGP parameters are $h = 0.665$, $\omega_m = 0.116$, $\omega_k=0$
		and $\omega_{rc} = 0.06$.  Errorbars represent expected accuracy 
        of the 200\,m telescope.}
	\end{figure}

	In the case that expansion history measurements are consistent with DGP,
	the question arises as to whether DGP could be distinguished from a 
	smooth dark energy model that had the same expansion history.  
	The additional information in linear perturbations as measured by weak
	lensing allows DGP to be distinguished even from a dark energy model 
	with an identical expansion history.  Figure \ref{f:results:DGPWL} shows
	the lensing spectra for a DGP cosmology similar to the best fit discussed 
	above, as well as the dark energy model with the same expansion history
    as in \cite{Fang:2008kc}.

	In principle one should consider the small amount of freedom within
	the DGP parameter set that could be used to make the DGP spectrum
	better fit the dark energy spectra.  However this is unlikely to 
	significantly change the spectrum as all relevant parameters are
	tightly constrained by the CMB and BAO. For example it is clear
	from Figure \ref{f:results:DGPWL} that the lensing spectra of the 
	two models would better agree if the amplitude of primordial scaler
	perturbations was increased in the DGP model. However, Planck 
	measurements would only allow of order half a percent increase while
	the disagreement is of order 10\%.  This is somewhat justified by the
	lack of correlations found in the \fr{} fisher matrices once all three 
	observational techniques are included.  In addition we have not 
	considered information from weak lensing in the nonlinear regime. 
	Adding nonlinear scales would only make our conclusion that DGP and 
    smooth dark energy are distinguishable with these observations more robust.

\section{Discussion}
	\label{s:discuss}

	We have shown that the first generation of 21\,cm intensity mapping instruments
	will be capable of constraining the HS \fr{} model (with $n=1$) down to a
	field value of $|f_{R0}| \lesssim 2 \times 10^{-5}$ at 95\% confidence 
    (Figure~\ref{f:results:FRlowres}).
	This is an 
	order of magnitude tighter than constraints currently available from 
	galaxy
	cluster abundance \cite{fRcluster}.  Furthermore, model parameters in
        this regime are not ruled out by Solar System tests. 

    In comparing Figures 
	\ref{f:results:FRfisher} and \ref{f:results:FRlowres} it is clear
	that a more advanced
	experiment, with resolution improved by a factor of two, would
	further half the allowed value of $|f_{R0}|$.  It should be noted however,
	that halving of the allowed parameter space does not correspond to a factor
	of four increase in information.  Deviations in the lensing spectrum scale
	sub-linearly in the \fr{} parameters, enhancing the narrowing of
	constraints as information is added (see Section \ref{s:results}).

While we have concentrated on a particular $f(R)$ model, many viable 
functional forms for $f(R)$ have been proposed in the literature 
\cite{Nojiri:2007as,Starobinsky:2007hu,Appleby:2007vb}.  The predictions for the
growth of structure in these different models agree qualitatively: the 
gravitational force is enhanced by 4/3 within $\lambda_C$, enhancing
the growth on small scales.  However, there are quantitative differences in 
the model predictions due to the different evolution of $\lambda_C$ over 
cosmic time.  Our results for the HS model with different values of $n$ should 
thus cover a range
of different functional forms for \fr{}.  Table~\ref{t:results:fr} shows
that our constraints do not depend very sensitively on the value of $n$.  
This is because the weak lensing measurements cover a wide range of scales
as well as redshifts.  Furthermore, it is straightforward
to map the enhancement in the linear $P(k,z)$ at given $k$ and $z$ from the HS model 
considered here to any other given model, to obtain approximate constraints
for that model.

Future cluster constraints will almost certainly improve on the current
limits of $|f_{R0}| \lesssim\mbox{few}\: 10^{-4}$ \cite{fRcluster}.  However, for smaller field values,
the main effect of $f(R)$ gravity shifts to lower mass halos, since the highest
mass halos are chameleon-screened (see Fig.~2 in \cite{HPMhalopaper}).  Hence, future 
cluster constraints will depend
on the ability to accurately measure the halo abundance at masses around
few $10^{14} M_{\odot}$ and less.  
Furthermore, the constraints from cluster abundances depend sensitively
on the knowledge of the cluster mass scale, and are already 
systematics-dominated \cite{fRcluster}.  Weak lensing constraints have a 
completely independent set of observational systematics, and are
in principle less sensitive to baryonic or astrophysical effects.  Thus,
the forecasted constraints on modified gravity presented here are quite 
complementary to constraints from cluster abundances.

	The processes that produce the BAO feature in the matter
	power spectrum are understood from first principles. In addition the BAO
	length scale can be extracted even in the presence of large 
	uncertainties in biases and
	mass calibrations.  Likewise, weak lensing on
	large scales is well understood, with baryonic physics being
	much less important than on smaller scales \cite{Zhan:2004wq}.
	In addition the dominant systematics present
	in optical weak lensing surveys are instrumental
	in nature and not intrinsic to the quantities being measured.
	While 21\,cm intensity mapping is as yet untested, instrumental
	systematics will be very different from those that affect the optical.
	
	In the case of this study, and more generally for cosmological models 
    which substantially modify structure
	formation, the motivation for higher resolution comes not from improved
	BAO measurements but from better weak lensing reconstruction.  
	Higher resolution not only makes weak lensing information available at
	higher multi-poles, but improves the accuracy at which lensing can
	be reconstructed on all scales.

	The inclusion of lensing information in the nonlinear regime
	was crucial, and largely responsible for the competitiveness of these 
	forecasts.  As seen in Figure \ref{f:obs:lensing}, much of the 
	constraints 
	come from multi-poles in the nonlinear regime.  It should be noted that
	for the higher resolution experiment considered, the minimum scale
	is limited not by the resolution at high redshift, but by the saturation
    of information in nonlinear source structures at low
	redshift \cite{Lu:2009je}. 
	
	Our constraints from lensing are conservative since only one wide source 
	redshift
	bin was considered, limited to $\l<600$ as described above.  To maximize 
	information, the  source redshift range could be split into multiple bins,
	properly considering the correlation in the lensing signal between them; 
    a process known as lensing tomography.
	The low redshift bin would be limited as above, and the high redshift
	bin would be limited by the resolution to $\l \approx 850$ at $z =  2.5$.
	However in intermediate bins, the lensing signal could be reliably
	reconstructed above $\l\approx1000$.

	Unlike most smooth dark energy models, such as quintessence, constraints
	on the models considered here are chiefly sensitive to structure formation,
	as is clear from Figure \ref{f:results:FRfisher}.
	No experiment in the foreseeable future will be able to improve upon 
	the constraints we project on these models using exclusively expansion 
	history probes 
	(except by breaking degeneracies).
	These forecasts show that 21\,cm intensity mapping is not only sensitive
	to a cosmology's expansion history through the BAO, but also to structure
	growth through weak lensing.  The weak lensing measurements cannot
	compete with
	far off space based surveys like Euclid or JDEM, which will have galaxy
	densities of order $100\,\mathrm{arcmin}^{-2}$\cite{Albrecht:2006um}
	and resolution to far greater $\l$.
	However, cylindrical 21\,cm experiments are realizable on a
	much shorter time scale and at a fraction of the cost.  
	In addition, the
	measurements considered here are approaching the limit at which \fr\
	models can be tested.  For $|f_{R0}|$ much less than $10^{-6}$ the 
	chameleon mechanism becomes important before there are observable 
	modifications to structure growth, reducing the motivation to further 
	study these models.
	
	It has also been shown that, for these experiments, a BAO measurement
	consistent with \LCDM{} would definitively rule out DGP without a 
	cosmological constant as a cosmological model.  Even in the case
	that a BAO measurement consistent with DGP is made, the model is still
	distinguishable from an exotic smooth dark energy model through
	structure growth.  The former result is not surprising given that DGP
	is now in conflict with current data \cite{Fang:2008kc}.  However
	it is illustrative that a single experiment can precisely probe both
	structure formation and expansion history.  Even a dark energy
	model that conspires to mimic DGP is, to a large extent, 
	distinguishable.
	
	We have studied the effects of modified gravity theories on observational
	quantities for future 21\,cm surveys.  Because these surveys measure the
	distribution of galaxies on large angular scales over large parts of the
	sky, they are well suited to measure the expected deviations relative
	to standard general relativity.  We have computed the predictions
	of modified gravity in the linear and nonlinear regimes, and compared
	to the sensitivity of future surveys.  We find that a large part of
	parameter space can be tested.

\begin{acknowledgments}

We would like to thank Tingting Lu for helpful discussions.
KM is supported by an NSERC Canadian Graduate Scholars-M scholarship.  
FS is supported by the Gordon and Betty Moore Foundation at Caltech.
PM acknowledges support of the Beatrice D. Tremaine Fellowship.

\end{acknowledgments}

	\bibliography{spires,fabian,/home/pmcdonal/Latex/cosmo}

%

\end{document}